\documentstyle[11pt,amssymb,epsf]{article}

\def\ben{\begin{equation}}
\def\een{\end{equation}}

  \let\n=\nu

\let\C=\Chi

\def\nn{\nonumber} \def\bd{\begin{document}} \def\ed{\end{document}}
\def\ds{\documentstyle} \let\fr=\frac \let\bl=\bigl \let\br=\bigr
\let\Br=\Bigr \let\Bl=\Bigl
\let\bm=\bibitem
\let\na=\nabla
\let\pa=\partial \let\ov=\overline
\newcommand{\be}{\begin{equation}}
\newcommand{\ee}{\end{equation}}
\def\ba{\begin{array}}
\def\ea{\end{array}}
\def\ft#1#2{{\textstyle{{\scriptstyle #1}\over {\scriptstyle #2}}}}
\def\fft#1#2{{#1 \over #2}}
\def\del{\partial}
\def\vp{\varphi}
\def\sst#1{{\scriptscriptstyle #1}}
\def\oneone{\rlap 1\mkern4mu{\rm l}}
\def\td{\tilde}
\def\wtd{\widetilde}
\def\ie{\rm i.e.\ }
\def\dalemb#1#2{{\vbox{\hrule height .#2pt
        \hbox{\vrule width.#2pt height#1pt \kern#1pt
                \vrule width.#2pt}
        \hrule height.#2pt}}}
\def\square{\mathord{\dalemb{6.8}{7}\hbox{\hskip1pt}}}
\newcommand{\ho}[1]{$\, ^{#1}$}
\newcommand{\hoch}[1]{$\, ^{#1}$}
\newcommand{\bea}{\begin{eqnarray}}
\newcommand{\eea}{\end{eqnarray}}
\newcommand{\ra}{\rightarrow}
\newcommand{\lra}{\longrightarrow}
\newcommand{\Lra}{\Leftrightarrow}
\newcommand{\bp}{\tilde \beta^\prime}
\newcommand{\tr}{{\rm tr} }
\newcommand{\Tr}{{\rm Tr} }
\def\0{{\sst{(0)}}}
\def\1{{\sst{(1)}}}
\def\2{{\sst{(2)}}}
\def\3{{\sst{(3)}}}
\def\4{{\sst{(4)}}}
\def\5{{\sst{(5)}}}
\def\6{{\sst{(6)}}}
\def\7{{\sst{(7)}}}
\def\8{{\sst{(8)}}}
\def\n{{\sst{(n)}}}
\def\cA{{{\cal A}}}
\def\cB{{{\cal B}}}
\def\cF{{{\cal F}}}
\def\cH{{{\cal H}}}
\def\tV{\widetilde V}
\def\tW{\widetilde W}
\def\tH{\widetilde H}
\def\tE{\widetilde E}
\def\tF{\widetilde F}
\def\tA{\widetilde A}
\def\im{{i}}
\def\tY{{{\wtd Y}}}
\def\ep{{\epsilon}}
\def\vep{{\varepsilon}}
\def\R{\rlap{\rm I}\mkern3mu{\rm R}}
\def\bD{{{\bar D}}}

\def\R{\rlap{\rm I}\mkern3mu{\rm R}}
\def\bD{{{\bar D}}}
\def\R{{{\Bbb R}}}
\def\C{{{\Bbb C}}}
\def\H{{{\Bbb H}}}
\def\CP{{{\Bbb C}{\Bbb P}}}
\def\RP{{{\Bbb R}{\Bbb P}}}
\def\Z{{{\Bbb Z}}}
\def\bA{{{\Bbb A}}}
\def\bB{{{\Bbb B}}}
\def\bC{{{\Bbb C}}}
\def\bD{{{\Bbb D}}}
\def\bE{{{\Bbb E}}}
\def\bZ{{{\Bbb Z}}}
\def\Re{{{\frak{Re}}}}
\def\Im{{{\frak{Im}}}}
\def\cosec{{\,\hbox{cosec}\,}}
\def\Gm{{\Gamma_{\!\! -}}}
\def\Gp{{\Gamma_{\!\! +}}}
\def\stan{{standard }}
\def\nonstan{{supernumerary }}

\newcommand{\tamphys}{\it Center for Theoretical Physics,
Texas A\&M University, College Station, TX 77843}

\newcommand{\upenn}{\it Department of Physics and Astronomy,\\ University
of Pennsylvania, Philadelphia, PA 19104}

\newcommand{\brussels}{\it Physique Th\'eorique et Math\'ematique,
Universit\'e Libre de Bruxelles,\\ Campus Plaine C.P. 231, B-1050
Bruxelles, Belgium}

\newcommand{\auth}{M. Cveti\v c\hoch{*1},  
H. L\"u\hoch{\dagger\ddagger2} and C.N. Pope\hoch{\dagger2}}

\begin{document}
\begin{flushright}
MIFP-04-13\ \ \ UPR-1084-T\ \ \ USTC-ICTS-04-16\\
{\bf hep-th/0407058}\\
July\  2004
\end{flushright}

\vspace{10pt}

\begin{center}

{\large {\bf  Charged Rotating Black Holes in 
Five Dimensional $U(1)^3$ Gauged ${\cal N}=2$ Supergravity}}

\vspace{20pt}
\auth

\vspace{20pt}{\hoch{*}\it Department of Physics and Astronomy,\\
University of Pennsylvania, Philadelphia, PA 19104, USA}

\vspace{10pt}{\hoch{\dagger}\it George P. \& Cynthia W. Mitchell
Institute for Fundamental Physics,\\ Texas A\& M University,
College Station, TX 77843-4242, USA}

\vspace{10pt} {\hoch{\ddagger}\it Interdisciplinary Center for
Theoretical Study, \\ University of Science \& Technology of China,
Hefei, Anhui 230026, China}

%
%
%

\vspace{40pt}

\underline{ABSTRACT}
\end{center}

    We obtain the general solution for non-extremal 3-charge dilatonic
rotating black holes in the $U(1)^3$ gauged five-dimensional ${\cal N}=2$ 
supergravity coupled to two vector multiplets, in the case where the
two rotation parameters are set equal.  These solutions encompass all
the previously-known extremal solutions, and, by setting the three
charges equal, the recently-obtained non-extremal solutions of ${\cal
N}=2$ gauged five-dimensional pure supergravity.

{\vfill\leftline{}\vfill \vskip 10pt \footnoterule {\footnotesize
\hoch{1} Research supported in part by DOE grant
DE-FG02-95ER40893, NSF grant INTO3-24081, and the\\
$\phantom{xxxxi}$  Fay R. and Eugene L.
Langberg Chair.}\vskip 2pt
{\footnotesize
\hoch{2} Research supported in part by DOE grant
DE-FG03-95ER40917.}
}

\pagebreak


    Valuable tests of the AdS/CFT correspondence can be performed by
taking charged black holes with non-zero cosmological constant as 
gravitational backgrounds \cite{mal,gkp}.  The black-hole charges
play the r\^ole of R-charges in the dual field theory \cite{cg}.
Furthermore, one can gain insight into the phase structure of the
strongly-coupled dual field theory by studying the thermodynamic 
stability of the black-hole solutions, and the analogue of the
Hawking-Page transition \cite{cg,cg2,cejm}.  The first examples
of non-extremal charged black holes in five dimensions, as solutions of 
a gauged supergravity theory, were obtained in \cite{bcs}.  These, 
and some higher-dimensional generalisations obtained in \cite{10a},
were all non-rotating.

    Charged rotating black holes in four dimensional theories
with a cosmological constant were obtained long ago \cite{carter}, but
until recently no analogous five-dimensional charged rotating solutions
were known, except in certain extremal BPS limiting cases 
\cite{klemm1,reall}.
In a recent paper \cite{cvlupo}, we constructed general solutions for
charged rotating black holes in five-dimensional gauged ${\cal N}=2$
pure supergravity, in the case where the two angular momenta are
taken to be equal.  These non-extremal solutions encompass the extremal
solutions of \cite{klemm1,reall} as special cases.   By instead
setting the charge to zero, the solutions in \cite{cvlupo} reduce to 
the rotating five-dimensional Kerr-de Sitter black holes of 
\cite{hawhuntay},
in the special case where the two rotation parameters are set equal.   
 
    In this letter we extend our previous results, by constructing a
general class of non-extremal charged rotating black hole solutions in
the five-dimensional $U(1)^3$ gauged theory of ${\cal N}=2$ supergravity
coupled to two vector multiplets.  We obtain the general non-extremal
solutions of this dilatonic theory, with three independent electric
charges, subject to the specialisation that the two angular momenta in
the orthogonal 4-space are set equal.  These 3-charge solutions are 
important for probing fully the microscopic degrees of freedom associated
with the three R-charges in the dual ${\cal N}=4$ CFT on the boundary,
without the loss of information that would be inherent if the three
charges were set equal. 

   Our new 3-charge solutions are generalisations to the gauged theory
of the 3-charge spinning black hole solutions (with two rotation
parameters set equal) of the corresponding five dimensional {\it
ungauged} supergravity, obtained in \cite{cvetyoum}.  They also, of
course, specialise to our previous results in \cite{cvlupo} if one
sets the three electric charges equal, under which circumstance the
two dilatonic scalars decouple and become constant.

    The bosonic sector of the five-dimensional ${\cal N}=2$ gauged
supergravity coupled to two vector multiplets is described by the 
Lagrangian
\be
e^{-1}\, {\cal L} = R - \ft12{\del\vec\varphi}^2 - 
  \ft14\sum_{i=1}^3 X_i^{-2}\, {(F^i)}^2 -\lambda\, 
  \sum_{i=1}^3 X_i^{-1} + \ft1{24} \ep_{ijk}\, \ep^{\mu\nu\rho\sigma\lambda}
  F^i_{\mu\nu}\, F^j_{\rho\sigma}\, A^k_{\lambda}\,,\label{d5lag}
\ee
where $\vec\varphi=(\varphi_1,\varphi_2)$, and
\be
X_1= e^{-\fft1{\sqrt6}\varphi_1 -\fft1{\sqrt2} \varphi_2}\,,\qquad
X_2= e^{-\fft1{\sqrt6}\varphi_1 +\fft1{\sqrt2} \varphi_2}\,,\qquad
X_3 = e^{\fft2{\sqrt6}\varphi_1}\,.
\ee
The gauge-coupling constant $g$ is related to $\lambda$ by
$\lambda=-g^2$.

   The solutions that we have obtained are as follows:
\bea
ds_5^2 &=& -\fft{Y-f_3}{R^2}\, dt^2 + \fft{r^2\, R}{Y}\, dr^2 +
   R\, d\Omega_3^2 + \fft{f_1 - R^3}{R^2}\, (\sin^2\theta d\phi + 
\cos^2\theta d\psi)^2\nn\\
&& - \fft{2 f_2}{R^2}\, dt\, (\sin^2\theta d\phi + \cos^2\theta d\psi)\,,
\label{metres}\\
A^i &=& \fft{\mu}{r^2\, H_i}\, \Big(s_i\, c_i\, dt + \ell\, (c_i\, s_j\, s_k
      - s_i \, c_j\, c_k)\, (\sin^2\theta d\phi + \cos^2\theta d\psi)
           \Big)\,,\label{adef}\\
X_i &=&  \fft{R}{r^2\, H_i}\,,\ \ \ i=1,2,3
\eea
where
\bea
R &\equiv & r^2\, (\prod_{i=1}^3 H_i)^{\fft{1}{3}}\,,\qquad
H_i \equiv 1 + \fft{\mu\, s_i^2}{r^2}\,,\nn\\
d\Omega_3^2 &=& d\theta^2 + \sin^2\theta d\phi^2 + \cos^2\theta d\psi^2\,,
\eea
and $s_i$ and $c_i$ are shorthand notations for
\be
s_i\equiv \sinh\delta_i\,,\qquad c_i \equiv \cosh\delta_i\,,\qquad
i=1,2,3\,.
\ee
Note that in the expressions (\ref{adef}) for the vector potentials
$A^i$, the triplet indices $(i,j,k)$ are all unequal: $(i\ne j\ne k\ne i)$.
The functions $(f_1, f_2,f_3, Y)$ are given by
\bea
f_1 &=& R^3 + \mu\, \ell^2\, r^2 + \mu^2\, \ell^2\, \Big[ 2\Big(\prod_i 
  c_i - \prod_i s_i\Big) \, \prod_j s_j - \sum_{i<j} s_i^2 \, s_j^2\Big]
\,,\nn\\
f_2 &=& \gamma\, \ell\, \lambda\, R^3 + \mu\, \ell\, \Big(\prod_i
  c_i - \prod_i s_i\Big)\, r^2 + \mu^2\, \ell\, \prod_i s_i\,,\nn\\
f_3 &=& \gamma^2\, \ell^2\, \lambda^2\, R^3 + \mu\, \ell^2\, \lambda\, 
\Big[ 2\gamma\, \Big(\prod_i c_i - \prod_i s_i\Big)-\Sigma\Big]\, r^2 
      \\
&&+ \mu\, \ell^2 - \lambda\, \Sigma\, \mu^2\, \ell^2\, \Big[
    2\Big(\prod_i c_i - \prod_i s_i\Big)\, \prod_j s_j - \sum_{i<j}\, 
s_i^2\, s_j^2\Big] + 2\lambda\, \gamma\, \mu^2\, \ell^2\, \prod_i s_i\,,\nn
\\
Y &=& f_3 -\lambda\, \Sigma\, R^3 + r^4 -\mu\, r^2\,,\nn
\eea
where
\be
\Sigma\equiv 1 + \gamma^2\, \ell^2\, \lambda\,.
\ee
It is helpful to note that $\sqrt{-g}$ takes a simple form, namely
$\sqrt{-g} = r\, R\, \sin\theta\, \cos\theta$.

   We arrived at the above solution by making conjectures for the
expressions for the metric, vector potentials and dilatonic scalars 
that reduced to previously-known cases under appropriate limits.  In
particular, we were guided by the results for the ungauged case in
\cite{cvetyoum}, and the results for the non-dilatonic 
gauged case (\ie with three equal charges) in \cite{cvlupo}.
Verifying that the conjectured configuration solves the equations of motion
following from (\ref{d5lag}) is then a straightforward mechanical exercise, 
which is most easily accomplished with the aid of a computer.  (We used
Mathematica for this purpose.)   
There are six free parameters in the solution, namely $(\mu, \delta_1,
\delta_2,\delta_3,\ell, \gamma)$.  The constant $\mu$, together with the
three ``non-extremality parameters''
$\delta_i$, characterise the mass and the three
electric charges associated with the three vector potentials $A^i$.  The
parameter $\ell$ characterises the rotation of the black hole.  One can
define ``physical'' mass, charge and angular momentum parameters $M$, 
$Q_i$ and $J$, according to
\bea
M &=& \ft12 \mu\, \sum_i(s_i^2 + c_i^2)\,,\nn\\
Q_i &=& \mu\, s_i\, c_i\,,\label{mqjdef}\\
J &=& \mu\, \ell\, \Big(\prod_i c_i -\prod_i s_i\Big)\,.\nn
\eea

   The 
sixth constant, $\gamma$,  is
the analogue of the extra parameter $\beta$ found in the charged rotating
black holes of the pure ${\cal N}=2$ gauged supergravity theory in 
\cite{cvlupo}.  It is in fact a trivial parameter (see \cite{cvlupo},
and, for a complete proof of its triviality, \cite{madros}).
It is nonetheless useful to retain the
redundant parameter $\gamma$, since it provides a convenient way to
consider various limits.

     In order to make the global structure of the metrics more 
apparent, it is convenient to rewrite the metric (\ref{metres})
in terms of left-invariant 1-forms $\sigma_i$ on $S^3$.  Defining
\bea
\sigma_1 &=& \cos\td \psi\, d\td \theta + \sin\td\psi\, \sin\td\theta
\,d\td\phi\,,\nn\\
\sigma_2 &=& -\sin\td \psi\, d\td \theta + \cos\td\psi\, \sin\td\theta
\,d\td\phi\,,\nn\\
\sigma_3&=& d\td\psi + \cos\td\theta\, d\td\phi\,,
\eea
where
\be
\psi -\phi = \td \phi\,,\qquad
\psi +\phi = \td \psi\,,\qquad \theta = \ft12 \td\theta\,,
\ee
we find that (\ref{metres}) can be rewritten as
\be
ds_5^2 = -\fft{R\, Y}{f_1}\, dt^2 + \fft{r^2\, R}{Y}\, dr^2 +
\ft14 R\, (\sigma_1^2 + \sigma_2^2) + \fft{f_1}{4R^2}\, (\sigma_3 
-\fft{2 f_2}{f_1}\, dt)^2\,,\label{met2}
\ee
whilst the vector potentials in (\ref{adef}) become
\be
A^i = \fft{\mu}{r^2\, H_i}\, \Big(s_i\, c_i\, dt +\ft12 \ell\, 
(c_i\, s_j\, s_k
      - s_i \, c_j\, c_k)\, \sigma_3
           \Big)\,.
\ee

\newpage
\medskip
\noindent{{\bf Reductions to Previously-known Solutions}}
\medskip

    Various limits of our new solutions reduce to previously-known
cases.  These include the non-extremal 3-charge spinning black
hole solutions of the ungauged theory in \cite{cvetyoum} (specialised to
the case of equal angular momenta); the non-extremal charged rotating
solutions of the pure ${\cal N}=2$ gauged theory found recently in
\cite{cvlupo}; the BPS 3-charge rotating solutions of Klemm and Sabra
\cite{KSII}; and the BPS 3-charge rotating solutions of Gutowski and
Reall \cite{GRII}.  In detail, these various cases arise as follows:

\begin{itemize}

\item  The ungauged limit (\ie $\lambda=0$) leads to the  special case 
of the
solutions of \cite{cvetyoum}  where one
sets the two angular momenta parameters equal, \ie $\ell_1=\ell_2=\ell$.
This limit is obtained  from our solution by
setting $\lambda=0$, replacing $\mu$ by $2m$,  
and by redefining $r^2\to r^2+\ell^2$.

\item The equal-charge limit, \ie setting 
$\delta_1=\delta_2=\delta_3\equiv \delta$,
reduces to the solution found in \cite{cvlupo}.  Note that the 
parameters
$\beta$ and $J$ in \cite{cvlupo} are related to $\gamma$  and $\ell$ of
the present paper by
 \be
 \beta=\gamma\, e^\delta\,, \qquad   J=\ell\, e^{-\delta}\,,
\ee
while the parameters $M$ and $Q$ in \cite{cvlupo} can be read off 
from (\ref{mqjdef}).  The radial variable in \cite{cvlupo} is given by
sending $(r^2 + \mu\, \sinh^2\delta) \to r^2$.

\item
The Klemm-Sabra solution \cite{KSII}, which has closed timelike curves, 
is a BPS limit of our solution, obtained by taking 
\bea
\mu\to 0\, ,&\ &\   \delta_i\to  -\infty\, , \nonumber \\
Q_i=  \ft14 \mu\, e^{-2\delta_i}\, , &\ &\ \ell=\alpha\, 
\sqrt{\mu}\,,
\eea
where the three charges $Q_i$ and the constant $\alpha$ are kept finite and
non-zero.   The black hole mass and angular momentum, defined in 
(\ref{mqjdef}), are then given by
\be
M=-(Q_1+Q_2+Q_3)\,,  \qquad  J=2 \alpha \sqrt{Q_1Q_2Q_3}\,.
\ee
Note that since $\ell\to 0$ and the $\gamma$ parameter appears only in
 a product  with $\ell$, the solution does not depend on $\gamma$.

\item 
The Gutowski-Reall solution \cite{GRII} is a regular BPS limit
of our solution,  obtained by taking
\bea
 \mu \to 0\, ,&\ &\  \delta_i \to  +\infty\, ,  \\
Q_i=\ft14 \mu\,  e^{2\delta_i}\, , &\ &\ 
\gamma = \sqrt{\mu}\,, \qquad
 \ell =\fft{1}{\sqrt{-\lambda\,\mu }}\,, \nonumber
\eea
where the charges $Q_i$ are kept finite and non-zero. 
Note  that  $\gamma$ goes to zero, while $\ell$ goes to infinity, when the
limit is taken.  (The solution remains finite, however.)
The black hole  mass and the angular momentum, following from 
(\ref{mqjdef}),  are:
\be
M= + (Q_1+Q_2+Q_3)\,,  \qquad  
J=\fft{Q_1Q_2+Q_1Q_3+Q_2Q_3}{2\sqrt{-\lambda}\,
\sqrt{Q_1Q_2Q_3}}\,.
\ee

\end{itemize}    

\bigskip
\noindent{{\bf Further Remarks}}
\medskip

   The metric (\ref{met2}) has horizons at values of the radial
coordinate where $R\, Y\, f_1^{-1}$ vanishes.  In order to avoid naked
singularities, the outer horizon at $r=r_H$ should lie outside
the curvature singularity at $R=0$, and thus we require that 
it occur at the largest positive root of $Y(r_H)=0$.  In order to
avoid having closed timelike curves (CTCs), $f_1$ should be positive
for all $r>r_H$.  A detailed analysis of the restrictions on the
parameters in order to obtain solutions free of naked singularities or
closed CTCs is quite involved, and we shall not present it here.  It 
is analogous to the one given in \cite{cvlupo} for the case where
all three charges are equal.  Clearly, there exist appropriate ranges 
of the parameters for which such ``regular'' black holes arise.

  On the horizon, the Killing vector 
\be
l = \fft{\del}{\del t} - \fft{2 f_2(r_H)}{f_1(r_H)}\, \fft{\del}{\del 
\td\psi}
\ee
becomes null, and thus $r=r_H$ corresponds to a Killing horizon.  The 
surface gravity $\kappa$, which is constant over the horizon, can be 
calculated from
\be
\kappa^2 = (\del_\mu K)\, (\del^\mu K)\Big|_{r=r_H}\,,
\ee
where $K=\sqrt{-l^\mu\, l_\mu}$, implying that
\be
\kappa = \Big|\fft{Y'(r_H)}{4 r_H^2\, f_1(r_H)}\Big|\,.
\ee
The area of the Killing horizon is given by
\be
A = 2\pi^2\, \sqrt{f_1(r_H)}\,.
\ee
The Hawking temperature and entropy are therefore given by
\be
T = \Big|\fft{Y'(r_H)}{8\pi\, r_H^2\, f_1(r_H)}\Big|\,,\qquad
S= \ft12\pi^2\, \sqrt{f_1(r_H)}\,.
\ee

\section*{Acknowledgment}

We are grateful to ZhiWei Chong, Gary Gibbons, Dietmar Klemm, Malcolm
Perry and Wafic Sabra for useful discussions.  M.C. and C.N.P. are
grateful to the Relativity Group in DAMTP, Cambridge and the CERN
Theory Division, and M.C. is grateful to the Durham Centre 
for Fundamental Physics, for hospitality.  H.L. is supported in part by grants
from the Chinese Academy of Sciences and from the NSF of China with
Grant No: 10245001, 90303002.

\end{document}